\documentstyle[preprint,aps,amsfonts,prd]{revtex}

\begin{document}

\tightenlines

\title {Quantum corrected geodesics}
    
\author{Diego A.\ R.\ Dalvit $^{1}$ \thanks{dalvit@lanl.gov} and
Francisco D.\ Mazzitelli $^{2}$ \thanks{fmazzi@df.uba.ar}}

\address{$^1$ Theoretical Astrophysics, MS B288, Los Alamos National
Laboratory, Los Alamos, NM 87545, USA}

\address{$^2$ 
Departamento de F\'\i sica {\it J. J. Giambiagi}, 
Facultad de Ciencias Exactas y Naturales\\ 
Universidad de Buenos Aires- Ciudad Universitaria, Pabell\' on I\\ 
1428 Buenos Aires, Argentina}

\maketitle

\begin{abstract}

We compute the graviton-induced corrections to the 
trajectory of a classical test
particle. We show that the motion of the test 
particle is governed by an effective
action given by the expectation value (with respect to the graviton 
state) of the 
classical action. We analyze the quantum corrected equations of motion for 
the test particle in two 
particular backgrounds: a Robertson Walker spacetime and a $2+1$ dimensional
spacetime with rotational symmetry.
In both cases we show that
the quantum corrected trajectory is not a geodesic of the background metric.

\end{abstract}


\section{INTRODUCTION}

A full quantum theory of gravity is still out of reach. However, in situations
where the spacetime curvature is well below Planck's curvature, it is possible 
to compute some quantum gravity effects.
Indeed, metric fluctuations can be quantized using standard methods. The non 
renormalizability of the resulting quantum field theory is not an 
impediment for
making meaningful quantum corrections. The key point is to consider general 
relativity as an effective field theory \cite{dono}.

Although the leading long distance quantum corrections 
are expected to be too small
in realistic situations, the analysis of general 
relativity as an effective field  theory
is of conceptual interest. Moreover, tiny but measurable quantum 
gravity effects could
show up when measuring the decoherence of wavepackets of a 
non relativistic particle
subjected to the gravitational potential \cite{perci}. On the other hand, 
recent 
speculations raise
the length scale relevant for quantum gravity effects from Planck length to 
a TeV scale \cite{cern}. In this situation, the effects 
of metric fluctuations 
could be easier
to observe.

In the context of effective field theories, it is in principle 
possible to compute 
an effective action and  effective field equations for the mean 
value of the spacetime 
metric. The effective field equations (known as semiclassical 
Einstein equations or
 backreaction equations) include the backreaction of quantum matter fields and
of the metric fluctuations on the spacetime metric. These 
equations should be the
starting point to investigate interesting physical problems like, for example,
the dynamical evolution of a black hole geometry taking 
into account the evaporation 
process.

The backreaction equations have been investigated by 
several authors in the last
twenty years or so \cite {huetal}. 
However, due to the complexity of the problem 
(and also
to the non renormalizability  of the theory) most works considered 
scalar or spinor
quantum matter fields, but the  graviton contribution was simply omitted. 

It is in general
stressed that the graviton effects should be similar to 
those of a couple of massless, 
minimally coupled scalar fields. 
While this is true at the level of the backreaction
equations, there is an important physical difference that has been pointed only
recently \cite{nos}. When metric fluctuations are taken into account, the 
background geometry (i.e. the metric that solves 
the backreaction equations), turns 
out to be non physical. The reason is the following: 
any classical or quantum device
used to measure the spacetime geometry will also feel the graviton 
fluctuations. As the coupling between the  device
and the metric is non linear, the device will not measure the background
geometry, which therefore
is not the relevant physical quantity to compute. 
In particular, in Ref. \cite{nos} we have shown that, 
working in the Newtonian approximation,
the trajectory of a classical test particle is not a geodesic of the
background metric.
Instead its motion is determined by  a quantum corrected 
equation
that takes into account its coupling to the gravitons. Moreover, while the 
backreaction equations and their solutions depend on the gauge fixing of the
gravitons, this dependence cancels out in the 
quantum corrected equation of motion for the test particle.

The aim of this paper is to analyze the effect of the gravitons on the 
motion of a test particle beyond the 
Newtonian  approximation. In order to avoid technical complications, we will
assume we know a solution to the backreaction equations, and will focus 
only on the 
departure of the test particle's equation of motion from the geodesic equation
of the background metric. Moreover, we will consider models where it is easy
to fix completely the gauge of the gravitons and quantize the theory by taking 
into account 
the remaining degrees of freedom.

The paper is organized as follows. In Section II we prove that the effective 
action that governs the motion of the test particle is the mean value of 
the classical action. In Section III we consider Robertson Walker universes. 
We first briefly describe how to quantize the metric fluctuations in terms
of massless scalar fields. Then we compute the quantum corrections to the 
geodesic equation and solve the quantum corrected equations of motion 
perturbatively. In particular, we find the graviton corrections to the 
cosmological redshift. In Section IV we consider three dimensional gravity
coupled to a Maxwell field. Following Ref. \cite{ash1}, 
we first show that this model
is exactly soluble: one can fully fix the gauge and show that the degrees of 
freedom reside in the Maxwell field. Then we compute the quantum corrected
equation of motion for the test particle. We show that, even in regions 
where the background metric is locally flat, the trajectory of the test
particle is not a straight line. Section V contains our final remarks.

\section{EFFECTIVE ACTION FOR A TEST PARTICLE}

In this Section we will show that, when quantum 
metric fluctuations are taken into
account, the effective action for the test particle is the mean value of
its classical action. This result is summarized in Eq.(\ref{seef}) below
(the reader may want to accept this as a 
reasonable assumption and skip this 
section).

Consider pure gravity described by Einstein-Hilbert action
\footnote{
Our metric has signature $(-+++)$ and the curvature tensor is defined as
${R}^{\mu}_{\, \cdot \, \nu\alpha\beta} = 
\partial_{\alpha} \Gamma^{\mu}_{\nu\beta} - \ldots$, 
${R}_{\alpha\beta} = {R}^{\mu}_{\, \cdot \, \alpha\mu\beta}$ and
${R}= {g}^{\alpha\beta} {R}_{\alpha\beta}$. We use units
$\hbar=c=1$.} 
$S_{\rm G}=(2/\kappa^2) \int d^4x \sqrt{-g} R$, where $\kappa^2=32 \pi G$,
and imagine that in addition we have some type of matter content described
by an action $S_{\rm M}$. 
The effect of quantum metric fluctuations can be analyzed
with the background field method, expanding the whole action 
$S_{\rm G}+S_{\rm M}$ around a background metric as
$g_{\mu\nu} \rightarrow g_{\mu\nu} + \kappa h_{\mu\nu}$, and integrating over
the graviton field $h_{\mu\nu}$ to get an effective action for the
background metric. 
In order to fix the gauge one chooses a gauge-fixing function 
$\chi^{\mu}[g,h]$, a gauge-fixing action $S_{\rm gf}[g,h]=
-(1/2) \int d^4x \sqrt{-g} \chi^{\mu} g_{\mu\nu} \chi^{\nu}$,
and the corresponding ghost action $S_{\rm gh}$.

Imagine that in addition we have a classical test particle that moves in 
the above background metric
and we wish to study the effects of metric fluctuations
on it. We couple gravity to the particle by means of the standard action
$S_m[x]=-m \int \sqrt{-g_{\mu\nu}} dx^{\mu} dx^{\nu}$, where $x^{\mu}$
denotes the path of this test particle. The complete effective action 
$S_{\rm eff}$ for
the background metric $g_{\mu\nu}$ and for the test particle $m$ is
obtained by integrating the whole action 
$S\equiv S_{\rm G}+S_{\rm M}+S_{\rm gf}+S_m +S_{\rm gh}$ over the graviton and
ghost fields.
To evaluate it in the one loop approximation we first expand $S$
up to second order in gravitons.
The second order term reads
\begin{equation}
S^{(2)}= \int d^4y \sqrt{-g} h_{\mu\nu} F^{\mu\nu\rho\sigma} h_{\rho\sigma} -
\int d^4y \sqrt{-g} h_{\mu\nu} m^{\mu\nu\rho\sigma}  h_{\rho\sigma}
\end{equation}
where  ${\hat F}\equiv F^{\mu\nu\rho\sigma}$ 
is a second order differential operator that depends on the
background metric, and $m^{\mu\nu\rho\sigma}$
 is a tensor depending on the position and velocity of the
test particle,
\begin{equation}
m^{\mu\nu\rho\sigma}(y)=\frac{m \kappa^2}{8} \int d\tau
\delta^4(y-x(\tau)) \dot{x}^{\mu}\dot{x}^{\nu}\dot{x}^{\rho}\dot{x}^{\sigma}
\end{equation}
There is also a second order term in 
ghost fields, that for gauge-fixing functions
linear in the metric fluctuations  decouple from the gravitons, and couple only
to the background metric.

The result of the path integral is the classical action
$S_{\rm clas}=S_{\rm G}+S_{\rm M} + S_{m}$ plus the 
sum of two functional determinants,
\begin{equation}
S_{\rm eff}=S_{\rm clas} + \frac{i}{2} {\rm Tr} \ln ({\hat F} - {\hat m})
 - i {\rm Tr} \ln {\hat G}
\label{efe}
\end{equation}
where ${\hat G}$ is also a second order differential operator that arises from 
integrating over ghosts. 
Once the effective action is evaluated, one can derive the
equations of motion for the background metric $g_{\mu\nu}$, the so called
semiclassical Einstein equations, i.e. 
$\delta S_{\rm eff}/\delta g_{\mu\nu}=0$. To solve these equations one can
discard all contributions coming from the test particle, as they are
vanishingly small. As they stand, these equations
(obtained from the standard {\it in-out} effective action)
are neither real nor causal. In order to get real and causal equations of 
motion for the background metric, the {\it in-in} effective action must
be evaluated \cite{ctp}. Alternatively, one can take twice the real and
causal part of the propagators in the {\it in-out} field equations.
In both ways one gets semiclassical Einstein equations
suitable for initial value problems. 

>From the effective action
given above one can also derive the quantum corrected equation of
motion for the test particle, i.e. 
$\delta S_{\rm eff}/\delta x^{\rho}=0$,
which will be our main concern in what follows. The same comments about 
reality
and causality apply to this equation of motion. In this paper
we will work with the usual  {\it in-out} effective action and 
use the adequate propagators in the quantum corrected equations. 

In general it is extremely
complicated, if not impossible, to work out the functional traces in
Eq.(\ref{efe}), so 
several approximation methods have been developed to deal with them. 
However, in this paper we will only focus on
the quantum effects of the coupling 
between the test particle and gravitons.
We can make use of the fact that the test particle has a small mass,
so we can expand Eq.(\ref{efe}) in powers
of $m$ and just keep the leading contribution. In this way we find that
the whole effective action reads
\begin{equation}
S_{\rm eff}[g_{\mu\nu},x]= S_{\rm clas} + \frac{i}{2} {\rm Tr} {\ln {\hat F}}
- i {\rm Tr} \ln {\hat G}
- \int d^4x \sqrt{-g} \langle h_{\mu\nu} m^{\mu\nu\rho\sigma} h_{\rho\sigma}
\rangle
\end{equation}
The expectation value 
is taken with respect to the
graviton state. The effective action for the test 
particle will be the sum of the classical term $S_m[x]$ and this last term, 
so that we conclude that in fact that effective action is the expectation
value of the classical one
\begin{equation}
S_{\rm eff}[x] = \langle S_m[x] \rangle
\label{seef}
\end{equation}
It is important to stress that due to the non linear nature of the coupling
between gravity and test particle, the effective lagrangian is not the same
as the classical lagrangian evaluated in the expectation value for the
particle's path. 

The calculation described so far preserves the covariance in the background 
metric $g_{\mu\nu}$. Alternatively, one can fully fix the gauge of the
quantum fluctuations of the geometry and quantize the remaining degrees
of freedom. As can be easily proved, the argument leading to Eq. (\ref{seef})
remains unchanged, since it relies only on the fact that the test particle 
mass is small.

\section{QUANTUM CORRECTIONS TO GEODESICS FOR FLAT ROBERTSON-WALKER METRICS}

\subsection{Non covariant quantization}

In this subsection we briefly review the non-covariant method of quantization
for flat Robertson-Walker universes.
The metrics we are dealing with are therefore of the form
$ds^2=- dt^2 + a^2(t) d{\bf x}^2$, where $a(t)$ is the expansion coefficient
. The action for the matter content in RW
metrics has the form
\begin{equation}
S_{\rm M} = \int d^4x  \sqrt{-g} 
\left[
\frac{1}{2} (\rho + p) u^{\mu} u^{\nu} g_{\mu\nu} +
\frac{1}{2} (\rho + 3 p) 
\right]
\end{equation}
where $u^{\mu}$, $\rho$ and $p$ and the fluid's four-velocity, density and
pressure respectively. The associated classical Einstein equations are
\begin{equation}
R_{\mu\nu}=-\frac{1}{2} \left( T_{\mu\nu} - \frac{1}{2} g_{\mu\nu}
T_{\lambda}^{\lambda} \right)
\end{equation}
where the classical energy-momentum tensor is 
$T_{\mu\nu}=(\rho+p) u_{\mu} u_{\nu} - p g_{\mu\nu}$.

There are different ways to quantize the theory. One is based on the
background field method, which was described above.
Here we follow another
quantization procedure that starts from the classical theory of perturbations
in RW, developed in \cite{Lif}. One considers perturbations such that
$\delta \rho=\delta p=\delta u^{\mu}=0$, and metric perturbations $h_{\mu\nu}$
that satisfy $u^{\mu} h_{\mu\nu}=0$, and further imposes the gauge
conditions $h^{\mu\nu}_{~~;\nu}=0$. Finally one ends up with only two
independent components of the metric, $h_{+}$ and $h_{\times}$, which can
be expressed in terms of the original components of $h_{\mu\nu}$, and
that correspond to the two polarizations of a gravitational wave. The above
conditions on the metric imply that $h_{0\mu}=0$ and a transversality condition
${\tilde \nabla}_j h^{ij}=0$, where ${\tilde \nabla}_j$ denotes the covariant
derivative with respect to the spatial part of the metric. Both components
$h_+$ and $h_{\times}$, and also $h_i^j$, verify
the field equation for a minimally coupled massless scalar
field in RW
\begin{equation}
\Box \phi = - a^{-3} \frac{\partial}{\partial t} 
\left(
a^3 \frac{\partial}{\partial t} \phi 
\right)
+ \nabla^2 \phi =0
\label{scalar}
\end{equation}

To quantize we use the non-covariant quantization procedure of
\cite{Ford77,BLH78}. First one writes the second order term of the
action $S_{\rm G}+S_{\rm M}$ in terms of the two
independent degrees of freedom of the field,  $h_+$ and $h_{\times}$
\begin{equation}
S^{(2)}_{\rm G+M} = \frac{1}{2} \int d^4x \sqrt{-g} [ 
\partial_{\mu} h_+(x) \partial^{\mu} h_+(x) +
\partial_{\mu} h_{\times}(x) \partial^{\mu} h_{\times}(x) ]
\end{equation}
and then imposes equal-time canonical conmutation relations for the two 
scalar fields $[h_a({\bf x},t),\Pi_b({\bf x'},t)]= i \delta_{ab} 
\delta({\bf x}-{\bf x'})$, where $a,b=+,\times$ and $\Pi_a$ is the 
canonical momentum conjugate to $h_a$. This 
quantization procedure is 
equivalent to that for the individual modes $h_i^j$.
Instead of using canonical quantization, one can also 
do path integrals. One expands the action in terms of the individual modes
$h_i^j$ (or in terms of $h_+$ and $h_{\times}$) and integrates over
them in order to get an effective action for the background metric.
For the one loop effective action one needs the 
second order term of the expansion of the action in terms of metric
perturbations, namely
$S^{(2)}_{\rm G+M}=1/2 \int d^4y \sqrt{-g} h^i_j \Box h_i^j$, where $\Box$
denotes the scalar D'Alambertian operator. Finally one has to evaluate the
functional determinant of this differential operator.

\subsection{Quantum corrected geodesic equation}

Having summed up how to quantize metric perturbations in
RW, let us see how such quantum metric fluctuations affect the motion of a
classical test particle. As described in the previous section, the effective
action for the test particle is the expectation value of the classical 
action, namely
\begin{equation}
S_{\rm eff}[x]= - m \int \sqrt{-g_{\mu\nu}(x) dx^{\mu} dx^{\nu}}
- \frac{m \kappa^2}{8} \int d\tau \langle h_{ij}(x) h_{lm}(x) \rangle
\dot{x}^i \dot{x}^j \dot{x}^l \dot{x}^m
\end{equation}
where the dot denotes the derivative with respect to $\tau$. 
The graviton
two-point function can be expressed in terms of the scalar two-point function
$\langle \phi(x) \phi(x') \rangle$ as
\begin{equation}
\langle h_{ij}(x) h_{lm}(x') \rangle = - \frac{1}{3} a^2(t) a^2(t')
\left( \delta_{ij} \delta_{lm} - \frac{3}{2} \delta_{il} \delta_{jm}
- \frac{3}{2} \delta_{im} \delta_{jl} \right) 
\langle \phi(x) \phi(x') \rangle
\end{equation}
We recall that in these expressions the metric $g_{\mu\nu}$ is the solution
to the semiclassical Einstein equations that follow from quantizing gravity
in a RW universe. In the following we will assume that these equations have 
been solved and that the quantum corrected expansion factor $a(t)$ has been
found.

The geodesic equation for the test particle follows from
$\delta S_{\rm eff}[x]/\delta x^{\rho}=0$. For the temporal component
we get
\begin{equation}
\frac{d^2t}{d \tau^2} + a(t) a'(t) \left( \frac{d \bf{x}}{d \tau} \right)^2
- \frac{\kappa^2}{8} 
\dot{x}^i \dot{x}^j \dot{x}^l \dot{x}^m \frac{\partial}{\partial t}
G_{ijlm}[x(t)]=0
\label{geotemp}
\end{equation}
where $a'(t)\equiv da/dt$ and $G_{ijlm}[x(t)]$ is the coincident limit of
the graviton two-point function, evaluated along the trayectory of the 
particle. For the $n$-th spatial component ($n=1,2,3$) we obtain
\begin{equation}
\frac{d}{d \tau} \left(
a^2(t) \frac{d x^n}{d \tau} - \frac{\kappa^2}{2} G_{ijkl}[x(t)] 
\delta^{in} \dot{x}^j \dot{x}^k \dot{x}^l
\right)=0
\label{geoesp}
\end{equation}

Now let us solve Eqs.(\ref{geotemp},\ref{geoesp}) for $d {\bf x}/d \tau$ and
$dt/d \tau$. From Eq.(\ref{geoesp}) we see that the expression in 
parenthesis
is conserved. These conserved three quantities reflect the spatial 
translational invariance of RW metric, which is preserved upon the quantization
procedure. Therefore
\begin{equation}
a^2(t) \frac{d x^n}{d \tau} - \frac{\kappa^2}{2} G_{ijkl}[x(t)] 
\delta^{in} \dot{x}^j \dot{x}^k \dot{x}^l = \alpha
\label{consesp}
\end{equation}
where $\alpha$ is a dimensionless constant 
that depends on the initial velocity of the 
particle. Plugging this identity into Eq.(\ref{geotemp}) we find
\footnote{
This equation also follows from the very definition of the proper time.
Indeed, from $1=(dt/d\tau)^2 - a^2(t) (d{\bf x}/d\tau)^2$ we easily get
Eq.(\ref{constem}).
}
\begin{equation}
\frac{dt}{d \tau} = \sqrt{
1 + a^{-2}(t) \sum_{n=1}^{3} \left( \alpha + \frac{\kappa^2}{2}
G_{ijkl}[x(t)] \delta^{in} \dot{x}^j \dot{x}^k \dot{x}^l \right)^2 
}
\label{constem}
\end{equation}
Now we solve Eqs.(\ref{consesp},\ref{constem}) perturbatively in terms
of the coupling between the test particle and gravitons. Let us assume that
the initial velocity of the test particle is in the $x=x^1$ direction. The
zeroth order approximation corresponds to neglecting the coupling between the
particle and gravitons, which results in
\begin{eqnarray}
\frac{d x}{d \tau} &=& \frac{\alpha}{a^2(t)} \\
\frac{dt}{d \tau} &=& \sqrt{1 + \alpha^2 a^{-2}(t)} 
\label{geoclas}
\end{eqnarray}
Note that the limiting case of a light ray (null limit) $dx/dt=1$ is
obtained when $\alpha \rightarrow \infty$.

When the coupling is taken into account, 
we see that the particle still moves
in the same $x$ direction, and we get
\begin{eqnarray}
\frac{dx}{d\tau} &=& \frac{\alpha}{a^2(t)} 
\left( 1 + \frac{\alpha^2 \kappa^2}{3 a^2(t)} \langle \phi^2(t) \rangle
\right) \\
\frac{dt}{d\tau} &=& \sqrt{1+\alpha^2 a^{-2}(t)} 
\left(1 + \frac{\alpha^2 \kappa^2}{3 a^4(t)} 
\frac{\langle \phi^2(t) \rangle}{1+\alpha^2 a^{-2}(t)} \right)
\end{eqnarray}
where we expressed the graviton two point function 
in terms of the scalar two-point function as
$G_{xxxx}(t)=(2/3) a^4(t) \langle \phi^2(t) \rangle$.
The speed of the particle results
\begin{equation}
\frac{dx}{dt}= \frac{\alpha a^{-2}(t)}{\sqrt{1+\alpha^2 a^{-2}(t)}}
\left( 1 + \frac{32}{3}  \pi G  \langle \phi^2(t) \rangle
\frac{\alpha^2 a^{-2}(t)}{1+\alpha^2 a^{-2}(t)} \right)
\label{qvel}
\end{equation}
This is the main result of this section. It expresses the quantum corrections
to the velocity of a test particle that moves in a flat Robertson-Walker
quantum background. In the null limit
\begin{equation} 
dx/dt \approx a^{-1}(t) [1+(32/3) \pi G \langle \phi^2(t) \rangle]
\end{equation}
describes the graviton correction to the cosmological redshift. 

To estimate the effect of this quantum correction on the classical trayectory
of the test particle, we first have to evaluate the two-point function
in the coincident limit, 
$\langle \phi^2(t) \rangle$. As is well known, this coincident limit
is divergent, so
a renormalization procedure is compelling. In the following we will calculate
$\langle \phi^2(t) \rangle$ for particular RW metrics, namely
$a(t)=a_0 e^{H t}$ (de Sitter) and $a(t)=a_0 t^c$

For de Sitter spacetime, the two-point function not only has UV problems
but also IR ones. However, in the late time limit $t=t' \gg H^{-1}$
it is possible to give an approximate form for the renormalized function.
It was shown by several authors 
\cite{Ford1,Ford2,Linde,Staro} that the coincident limit grows linearly
with the coordinate time, $\langle \phi^2(t) \rangle \approx H^3 t/2 \pi^2$.
Using that $\kappa^2 \propto R^{-1}_{\rm Planck}$ and that for de Sitter the
curvature is constant $R \propto H^2$, we conclude that the quantum 
correction is proportional to $(R/R_{\rm Planck}) F(t)$, where the function
$F(t)=\alpha^2 H t a_0^{-2} e^{-2 H t}/(1+\alpha^2 a_0^{-2} \exp^{-2 H t})$ 
decreases exponentially for late times. The velocity of the test particle in 
the late time limit is therefore given by
\begin{equation}
\frac{dx}{dt} = \frac{\alpha a_0^{-2} e^{-2H t}}
{\sqrt{1+\alpha^2 a_0^{-2} e^{-2 H t}}} 
\left( 1+ \frac{16 G \hbar H^2 F(t)}{3 \pi c^5}  \right)
\label{vel1}
\end{equation}
where we have restored units $\hbar$ and $c$.

As we pointed out before, the scale factor $a(t)$ should be a solution to 
the semiclassical Einstein equations. A perturbative solution will be of the
form $a(t)=a_{\rm clas}(t) + \delta a (t)$, $a_{\rm clas} $ 
being the classical scale
factor and $\delta a \ll a_{\rm clas}$. It is well known that the semiclassical
Einstein equations admit de Sitter solutions \cite{wada} $a(t)=a_0 e^{Ht}$
with 
$H=H_{\rm clas}(1+\gamma {H_{\rm clas}^2\over R_{\rm Planck}}), \,\,\gamma =
O(1)$. Therefore, as long as 
${H_{\rm clas}^3 t\over R_{\rm Planck}}\ll 1$
the correction to the scale factor is given by ${\delta a\over a_{\rm clas}}
\simeq \gamma {H_{\rm clas}^3 t\over R_{\rm Planck}}$. 
Replacing $a(t)=a_{\rm clas}(t) + \delta 
a (t)$ in Eq. (\ref{vel1}) we obtain, to first order in all quantum corrections
\begin{equation}
\frac{dx}{dt} = \frac{\alpha a_0^{-2} e^{-2H_{\rm clas} t}}
{\sqrt{1+\alpha^2 a_0^{-2} e^{-2 H_{\rm clas} t}}} 
\left( 1+ \frac{16 G \hbar H^2 F(t)}{3 \pi c^5} 
-\gamma  {2+\alpha^2a^{-2}_{\rm clas}(t)\over 1+\alpha^2a^{-2}_{\rm clas}(t)}
 {H_{\rm clas}^3 t\over R_{\rm Planck}}\right)
\label{vel2}
\end{equation}
where $F(t)$ is to be evaluated with the classical value for the Hubble
parameter. 
This shows that the quantum correction to the geodesics coming from the 
graviton coupling (second term in Eq. (\ref{vel2})) and the one coming from 
the semiclassical
Einstein equations (third term) are of the same order of magnitude.

Consider now metrics with $a(t)=a_0 t^c$. Although these are not solutions
to the semiclassical Einstein equation, they are useful to illustrate
the corrections to the geodesics. 
In this case there are no infrared divergencies. In the
Appendix we give some details as to how to evaluate the renormalized
two-point function. The result is 
$\langle \phi^2(t) \rangle \propto t^{-2} \log(t^2 \mu^2)$, where $\mu$
is an (arbitrary) renormalization scale. Since for these metrics the curvature
is $R \propto t^{-2}$, we obtain that the quantum correction also has the
form $(R/R_{\rm Planck}) F(t)$, where now 
$F(t)=\alpha^2 a_0^{-2} t^{-2c} \log(t^2 \mu^2)/(1+\alpha^2 a_0^{-2} t^{-2c})$, which also decreases for long times. The velocity of the test particle is 
\begin{equation}
\frac{dx}{dt} = \frac{\alpha a_0^{-2} t^{-2c}}
{\sqrt{1+\alpha^2 a_0^{-2} t^{-2c}}} 
\left( 1+ \frac{2 c(2c-1) G \hbar F(t) }{3 \pi c^5 t^2} \right)
\end{equation}

\section{QUANTUM CORRECTIONS TO GEODESICS IN THREE DIMENSIONAL GRAVITY}

\subsection{Three dimensional General Relativity}

In this section we will consider 2+1 gravity coupled to Maxwell fields.
Under the assumption of rotational symmetry, this model is exactly
soluble. Moreover, it is possible to associate a well defined
quantum operator to the spacetime metric. Therefore, it is
particularly useful to analyze the effective action for a test
particle and the corrections to the geodesics. In this subsection we will
follow closely Refs. \cite{ash1,ash2}.

At the classical level, the theory is governed by the Einstein-Maxwell
equations, which read
\begin{eqnarray}
&& R_{ab}= 8\pi G\nabla_a\phi\nabla_b\phi\\
&&g^{ab}\nabla_a\nabla_b\phi = 0
\label{mee}
\end{eqnarray}
where the electromagnetic field has been written in terms of a
scalar field as $F_{ab}=\epsilon_{abc}\nabla^c\phi$. Assuming
rotational symmetry, the above equations can be easily solved. 
The metric can be written as
\begin{equation}
g_{ab}dx^a dx^b = e^{G\Gamma (r,t)}[-dt^2+dr^2]+r^2 d\theta^2
\end{equation}
Moreover,  the scalar field decouples from
the metric
\begin{equation}
g^{ab}\nabla_a\nabla_b\phi = 0 \rightarrow 
(-\partial_t^2+\partial_r^2)\phi = 0
\end{equation}
Therefore, one can solve the $1+1$ Klein Gordon equation for
$\phi$ and then determine $\Gamma$ from the Einstein equation.
The result is
\begin{equation}
\Gamma (r,t)={1\over 2}\int_0^r dr'~~r'~~[(\partial_t\phi)^2 +
(\partial_{r'}\phi)^2]
\label{ene}
\end{equation}
Note that, as $r\rightarrow\infty$, 
$\Gamma$ tends to a constant value $\Gamma (\infty ,t)
=H_0$. The metric becomes locally flat with a deficit angle
$2\pi (1- e^{-G H_0 /2})$.

To quantize the theory, one can promote $\phi$ to an operator
$\hat\phi$ describing a free quantum scalar field in $1+1$
dimensions. The spacetime metric is a secondary operator
that can be expressed in terms of $\hat\phi$ as
\begin{equation}
\hat{g}_{rr}=-\hat{g}_{tt}=e^{G\hat\Gamma}
\end{equation}
where $\hat\Gamma$ is the operator defined by Eq.(\ref{ene}) with
$\phi\rightarrow\hat\phi$.

For simplicity in what follows we will consider the metric
operator in the asymptotic region $r\rightarrow\infty$, where
the operator $\hat\Gamma$ is time independent. For a given
coherent state of the scalar field (denoted by $|F\rangle$ and 
peaked around a classical configuration $F(r,t)$), it
is easy to show that
\begin{eqnarray}
\langle F|\hat\phi |F \rangle &=& F(r,t)\nonumber\\
\langle F|\hat g_{rr} |F \rangle &=& \exp 
\left[ {1\over \hbar} \int_0^{\infty}  dw~~|F(w)|^2 
(e^{G\hbar w} - 1) \right]
\end{eqnarray}

For sufficiently low frequencies (i.e. when the Fourier transform
of the classical configuration is peaked around a low frequency),
the mean value of the metric operator can be approximated by
\begin{equation}
<F|\hat g_{rr} |F> = g_{rr} 
\left( 1 + \hbar {G^2\over 2} \int_0^{\infty} dw~w^2~|F(w)|^2 
\right)
\end{equation}
The first term is the value of the metric we would obtain from the
classical field equations for a classical scalar field 
configuration given by
$F(r,t)$. The second term represents a small quantum correction.
As in the classical case, for $r\rightarrow\infty$
the mean value of the metric describes
a locally flat spacetime, but with a quantum corrected deficit
angle.

\bigskip

\subsection{Effective action for a test particle}

According to our general discussion in Section 2, the effective action
for a test particle moving in the $2+1$ dimensional spacetime is
given by
\begin{equation}
S_{\rm eff}[x]=<S_m[x]>=-m 
<\int dt \sqrt{e^{G\hat\Gamma}(1-\dot r^2)-r^2\dot \theta ^2} >
\end{equation}
where the mean value is taken with respect to the coherent 
state $|F \rangle$. Here a dot denotes derivative with respect to
$t$. As in the previous section we will consider only the
asymptotic region where the metric operator is time 
independent.

We write the metric operator as 
$e^{G\hat\Gamma}= \langle e^{G\hat\Gamma} \rangle +\hat\Delta$. 
The effective lagrangian then becomes
\begin{equation}
L_{\rm eff}= -m \bar L \langle \sqrt{
\left[ 1+{\hat\Delta (1-\dot r^2)\over {\bar L}^2} \right] } \rangle 
\end{equation}
where $\bar L$ is proportional to the classical lagrangian evaluated in 
the mean value of the metric
\begin{equation}
\bar L =  \sqrt { \langle e^{G\hat\Gamma} \rangle (1-\dot r^2)-r^2\dot
\theta ^2 }
\end{equation}
Note that, after a redefinition of the angular variable $\theta\rightarrow
\sqrt{\langle  e^{G\hat\Gamma} \rangle} \theta$, $\bar L$ becomes 
proportional to the lagrangian of the test particle in a 
locally flat spacetime. The deficit angle is given by $2\pi (1- \sqrt{\langle  
e^{G\hat\Gamma} \rangle})$.

Assuming that the quantum fluctuations around the mean value are 
small \footnote {This is not always the case. See Ref. \cite{ash2}.}
we get
\begin{equation}
L_{\rm eff}= -m\bar L \left[ 1-{1\over 8}{(1-\dot r^2)^2\Delta^2\over
\bar L^4} \right]
\label{leff}
\end{equation}
where $\Delta^2=\langle \hat\Delta^2 \rangle =
 \langle (e^{G\hat\Gamma}- \langle e^{G\hat\Gamma} \rangle )^2\rangle$.
The above equation is the starting point to describe the quantum
corrections to the trajectory of the test particle.

Let us first consider a non relativistic motion of the particle.
In this situation we have
\begin{equation}
\bar L\simeq 
\sqrt{ \langle e^{G\hat\Gamma} \rangle} 
\left[ 1-{\dot r^2\over 2}-{r^2\dot\theta^2
\over 2 \langle e^{G\hat\Gamma} \rangle } \right]
\end{equation}
Therefore, the effective lagrangian can be approximated by
\begin{equation}
L_{\rm eff}\simeq -m
\sqrt{ \langle e^{G\hat\Gamma} \rangle }
\left[ 1- {1\over 8}\left ({\Delta g\over g}\right )^2\right ]
\left [ 1-{\dot r^2\over 2}-{r^2\dot\theta^2
\over 2 \langle e^{G\hat\Gamma} \rangle } 
\left( 1+ {1\over 2}\left ({\Delta g\over g}\right )
^2 \right) \right ] 
\label{lefnr}
\end{equation}
where $\left ({\Delta g\over g}\right 
)^2={\Delta^2\over  \langle e^{G\hat\Gamma} \rangle^2}$.

We can see from the Eq. (\ref{lefnr}) that in this 
nonrelativistic limit the 
effective lagrangian has, up to an irrelevant constant factor, 
the same form that $\bar L$,
but with a different deficit angle. Indeed, 
after the redefinition of the 
angular variable
$\theta \rightarrow \sqrt{\langle e^{G\hat\Gamma}\rangle}
\left( 1- {1\over 4}\left ({\Delta g\over g}\right )^2 
\right)\theta$,
the effective lagrangian becomes proportional
to the 
flat spacetime lagrangian. 
Therefore the trajectories will be
straight lines in a locally flat spacetime. However, the global
properties of the trajectories will be different from the ones
obtained with the mean value of the metric 
$\langle e^{G\hat\Gamma} \rangle$, since the deficit angle for the effective
lagrangian is now given by $2\pi (1- \sqrt{\langle e^{G\hat\Gamma}\rangle}
\left( 1- {1\over 4}\left ({\Delta g\over g}\right )^2 
\right) )$.

In the general case (a relativistic particle), the situation is
different. Indeed, one can prove that it is not possible to redefine
$\theta$ in order to bring $L_{\rm eff}$ (Eq. (\ref{leff}))
 to a flat spacetime form.
As a consequence, although the mean value of the metric is
locally flat, the test particle ``sees'' a much more
complex geometry.

The conclusion of this section is that, again, the trajectories
of the test particle do not coincide with the geodesics of
the mean value of the metric.

\section{FINAL REMARKS}

Let us summarize the new results contained in this paper. 
We have computed the 
quantum corrections to the trajectory of a test particle 
by taking into account
the quantum fluctuations of the spacetime metric. 
We have analyzed
two particular models where it is easy to fix completely the 
gauge of the quantum fluctuations and quantize the 
remaining degrees of freedom.

For a Robertson Walker spacetime, the fluctuations of the 
metric 
can be described by
two massless, minimally coupled scalar fields. The quantum corrected 
trajectory
has the same symmetries as the classical trajectory. 
However, it
contains a quantum correction
proportional to the graviton two point function and to the
initial velocity of the test particle. This additional 
term
produces, in particular,  a quantum correction to the gravitational 
redshift.

Let us assume that we solve the backreaction equations
perturbatively and find a solution $a(t)=a_c(t)+\delta a(t)$, where
$a_c(t)$ is the classical scale factor. Had we neglected the coupling
between gravitons and test particle, we would have concluded that the
test particle's trajectories coincide with the geodesics of the metric
$a(t)=a_c(t)+\delta a(t)$. However, this coupling induces an additional
correction to the equation of motion that is of the same order of
magnitude as the one produced by $\delta a(t)$.
(we have shown this in the particular case of a de Sitter solution and,
in a previous paper \cite{nos}, in the Newtonian approximation).
As a consequence, it is meaningless to compute $\delta a(t)$ and 
neglect the graviton effects on the motion of the particle, which
is the physical observable.

An interesting feature of our result is that the 
quantum corrections to the 
geodesic depend on the velocity of the test particle in such 
a way that one cannot
define an "effective metric" for the trajectory, i.e. 
a metric such that its
geodesics coincide with {\it all} the quantum
corrected trajectories.  
It is worth to note that, if one tries to define observationally 
an "effective spacetime curvature"
through a geodesic deviation equation, this effective curvature will
 be dependent on the initial four velocity of the geodesics under 
consideration.

In  the case of three dimensional general relativity, 
there are no propagating 
degrees of freedom associated to the geometry. At the classical level
one can make the degrees of freedom to reside in the matter field. 
At the quantum
level, the operator associated to the metric can be written 
in terms of the matter 
field operator.

In this model, given a quantum state of the matter fields, 
it is easy to compute the 
mean value of the metric and of any function of it. In particular, 
we computed
the mean value of the lagrangian for a test particle. We have shown that, 
even in
the asymptotic region, where both the classical metric and the mean 
value of the quantum metric
operator describe locally flat spacetimes, the test particle 
"feels" the quantum
fluctuations and the trajectory 
is not a straight line.

Now we would like to comment about related works. To our knowledge, 
the fact that the mean value of the metric is not enough to describe 
the spacetime geometry 
when the graviton contribution is taken into account,
was first pointed out  in Ref. \cite{vilkocqg}.
It was stressed there that one can assign an effective metric to a 
given observable ${\cal O} (g_{\mu\nu})$, through the identity
\begin{equation}
g^{\rm eff}_{\mu\nu}= {\cal O}^{-1}\langle {\cal O} (g_{\mu\nu})\rangle
\end{equation}
The effective metric obviously depends on $\cal O$.
We agree with this point of view. Indeed, from our results it is
easy to illustrate this fact. Consider for example the quantum 
corrected velocity of the test particle given in Eq. (\ref{qvel}).
Taking into account the classical result for the velocity,
one can introduce an "effective scale factor" through the 
identity
\begin{equation}
 \frac{\alpha a^{-2}_{\rm eff}(t)}{\sqrt{1+\alpha^2 a^{-2}_{\rm eff}(t)}}
= \frac{\alpha a^{-2}(t)}{\sqrt{1+\alpha^2 a^{-2}(t)}}
\left( 1 + \frac{32}{3}  \pi G  \langle \phi^2(t) \rangle
\frac{\alpha^2 a^{-2}(t)}{1+\alpha^2 a^{-2}(t)} \right)
\label{qvel2}
\end{equation}
This gives ${a_{\rm eff}\over a}\simeq 1-{\alpha^2+a^2\over \alpha^2+2 a^2}
\frac{32}{3}  \pi G  \langle \phi^2(t) \rangle
\frac{\alpha^2 a^{-2}(t)}{1+\alpha^2 a^{-2}(t)}$. The "effective
scale factor" depends on the initial velocity of the particle.

In Ref.\cite{fordsvai} the authors  analyzed the graviton induced fluctuations
of horizons in Robertson Walker and Schwarszchild spacetimes. The analysis was
based on the study of the effects of gravitons on (nearly) null geodesics. 
They 
pointed out that, due to the interaction with the fluctuations of the metric,
there are two effects on the trajectories of photons: the mean geodesic 
will deviate
from the classical geodesic, and there will be stochastic fluctuations
around the mean value. They studied the stochastic fluctuations and 
neglected the deviation of the mean value. 
In this sense, our work is complementary to Ref. \cite{fordsvai},
since we computed the mean value corrections. In our framework, 
the stochastic fluctuations could be analyzed by
using the CTP formalism to compute the effective action for the test
particle. It can be shown that the imaginary part of this CTP effective 
action
introduces a noise term in the 
equation of motion (similar ideas have been applied to the 
semiclassical
Einstein equations, see for example \cite{varios}).

In this paper we fixed completely the gauge of the metric
fluctuations  before quantization. Alternatively, one could use 
the covariant method described in Section II. We showed in a 
previous work \cite{nos} that the solution to the backreaction 
equation and the quantum corrections to the geodesics are both
dependent on the gauge fixing procedure. In the Newtonian approximation,
this dependence cancels when computing the trajectory of the test
particle. Whether this is true or not beyond the Newtonian approximation
 is an open question, that
will be addressed in a  forthcoming paper.

\section{Acknowledgments}
We acknowledge the support  
from Universidad de Buenos Aires, Fundaci\'on Antorchas and 
CONICET (Argentina).
We would like to thank M. Banados and D. Tiglio for 
pointing out Refs. \cite{ash1,ash2}
to us.

\newpage
\appendix
\section{}

In this appendix we calculate the renormalized two-point function
$\langle \phi(x) \phi(x') \rangle$ in the coincide limit $x' \rightarrow x$
for a massless minimally coupled scalar field in flat Roberton-Walker
metrics with $a(t)=a_0 t^c$. Throughout this appendix we work in conformal 
time, $\eta=[a_0 (1-c)]^{-1} t^{1-c}$. The metric reads $ds^2=C(\eta)
(-d\eta^2 + d{\bf x}^2)$ where 
$C(\eta)=a^2(t)=a_0^{2/(1-c)} (1-c)^{2c/(1-c)} \eta^{2c/(1-c)}$.

The two-point function we wish to evaluate is basically the 
Hadamard function $D^{(1)}=\langle \{ \phi(x),\phi(x') \} \rangle$.
By means of the point-splitting technique, we separate the points $x,x'$
only in their temporal component 
$\Delta \eta \equiv \eta-\eta'=\epsilon \rightarrow 0$. 
The Hadamard function then takes the form \cite{BD}
\begin{eqnarray}
D^{(1)}(x,x') &=& - 
\frac{C^{-1/2}(\eta) C^{-1/2}(\eta')}{2 \pi^2\Delta \eta^2 } -
\frac{R}{24 \pi^2}
\left[ \frac{1}{2}  \log \left| \frac{\epsilon^2}{C \eta^2} \right| +
\gamma + \frac{1}{2} \psi(\frac{3}{2}+\nu) + \frac{1}{2} \psi(\frac{3}{2}-\nu) 
\right] + \nonumber \\
&& \frac{R}{48 \pi^2} + {\cal O}(\epsilon^2)
\end{eqnarray}
where $\nu=|1-3c|/(2|1-c|)$, $\gamma$ is Euler's constant and $\psi$ is
Euler's function. The first term on the right is the expression for $D^{(1)}$
in the conformally coupled case, which can also be expanded in powers of 
$\epsilon$
\begin{equation}
- \frac{C^{-1/2}(\eta) C^{-1/2}(\eta')}{2 \pi^2\Delta \eta^2 } =
-\frac{1}{8 \pi^2 \epsilon^2 \Sigma} + \frac{1}{24 \pi^2}
\left[ R_{\alpha\beta} \frac{t^{\alpha}t^{\beta}}{\Sigma} -\frac{1}{6} R
\right] + {\cal O}(\epsilon^2)
\end{equation}
where $t^{\mu}$ is a unit vector that parametrizes the direction of
splitting and $\Sigma=t_{\mu} t^{\mu}$.

To renormalize we substract the second order adiabatic expansion for
the Hadamard function, namely
\begin{equation}
D^{(1)}_{\rm ad}(\eta,\eta') =
-\frac{1}{8 \pi^2 \epsilon^2 \Sigma} -
\frac{R}{24 \pi^2} \left[ \frac{1}{2} \log(\mu^2 \epsilon^2)+ \gamma\right]
+ \frac{1}{24 \pi^2} R_{\alpha\beta} \frac{t^{\alpha}t^{\beta}}{\Sigma}
+ {\cal O}(\epsilon^2)
\end{equation}
where $\mu$ is an arbitrary scale with dimensions of energy. Finally
\begin{equation}
D^{(1)}_{\rm ren}(x,x) = \lim_{\epsilon \rightarrow 0} 
\left( D^{(1)}(\eta,\eta') - D^{(1)}_{\rm ad}(\eta,\eta') \right)
= \frac{R}{48 \pi^2} \log(C \mu^2\eta^2) 
\end{equation}
all constants having been absorbed into a redefinition of $\mu$. We can now
go back to coordinate time, and on using that for these metrics the scalar
curvature is $R=6 c (2c-1) t^{-2}$, we get the final result
$\langle \phi^2(t) \rangle =\frac{6 c (2c-1)}{96 \pi^2} 
t^{-2} \log(t^2 \mu^2)$.


\end{document}